# FQAM-FBMC Design and Its Application to Machine Type Communication

Yinan Qi, Milos Tesanovic
Samsung Electronics R&D Institute UK, Staines, Middlesex TW18 4QE, UK
{yinan.qi, m.tesanovic}@samsung.com

*Abstract*—In this paper, we propose a novel waveform design which efficiently combines two air interface components: Frequency and Quadrature-Amplitude Modulation (FQAM) and Filter Bank Multicarrier (FBMC). The proposed approach takes the unique characteristics of FQAM into consideration and exploits the design of prototype filters for FBMC to effectively avoid self-interference between adjacent subcarriers in the complex domain, thus providing improved performance compared with conventional solutions in terms of self-interference, spectrum confinement and complexity with negligible rate loss. The resulting waveform properties are proven in this paper to be particularly suitable for Machine Type Communications (MTC) devices due to the observed reduced PAPR and lowered energy consumption. MTC has created a new eco-system that gives rise to a plethora of interesting applications and new business opportunities in the fifth-Generation (5G) mobile system and services – the enabling technology for the emerging paradigm of Internet of Things (IoT).

## I. INTRODUCTION

Within the pivotal objective of developing the overall 5G RAN design, air interface design will play a crucial part since its components directly impact spectrum efficiency, out-of-band emissions, throughput and transmission delay, system capacity and reliability, as well as hardware complexity and energy efficiency [1]-[2]. With ambitious goals set for 5G of supporting services with different (and often diverging) requirements, a highly flexible 5G air interface design will be required to answer this demand. As MTC has been one of key driving forces to 5G, spectrally efficient support for heterogeneous services that have quite different requirements is becoming ever so important. Accordingly, several enabling air-interface design methods have been actively investigated to support flexible spectrum sharing.

Machine Type Communication (MTC), also commonly known as Machine-to-Machine (M2M), refers to exchange of information (whether wireless or wired) to and from machines without human involvement [3]-[4]. Two important components of air interface design for MTC devices are waveform and modulation schemes. The choice of waveforms plays a major role in the successful realization of 5G. Cyclic Prefix (CP)-OFDM, as a waveform widely used in current wired/wireless communication systems, has been considered as one of the candidates for 5G waveforms. CP-OFDM achieves perfect orthogonality of subcarrier signals in the complex domain that allows trivial generation of transmit signal through IFFT, trivial separation of the transmitted data symbols at the receiver through FFT, and trivial adoption to MIMO channels [5]-[8]. It is well-localized in time domain and thus suitable for delay critical applications and Time Division Duplexing (TDD) based deployments. However, the spectral efficiency of CP-OFDM is reduced because of the usage of CP and it has comparatively high out of band (OOB) leakage.

In this regard, FBMC is proposed as an alternative [9]-[10]. FBMC enhances the fundamental spectral efficiency because of the well-localized time/frequency traits adopted from a pulse shaping filter per subcarrier, thus reducing the overhead of guard band required to fit in the given spectrum bandwidth, while meeting the spectrum mask requirement. Furthermore, the effectively increased symbol duration is suitable for handling the multi-path fading channels even without CP overhead. Consequently, the FBMC system can reduce the inherent overheads such as CP and guard-bands in CP-OFDM. FBMC is attractive in specific asynchronous scenarios, where Coordinated Multi-Point Transmission and Reception (CoMP) and Dynamic Spectrum Access (DSA) in a fragmented spectrum are employed to support the much higher traffic demand in 5G.

However, to maintain the transmission symbol rate, the conventional FBMC system generally doubles the lattice density either in time or in frequency compared with OFDM while adopting OQAM [11]. In OQAM, in-phase and quadrature-phase modulation symbols are mapped separately with half symbol duration offset. Thus, OQAM-FBMC or SMT causes intrinsic interference that makes it not straightforward to apply conventional pilot designs and corresponding channel estimation algorithms as well as MIMO schemes as in CP-OFDM systems. In this regard, QAM-FBMC system which can transmit the QAM symbols is proposed to enable fundamental spectral efficiency enhancement whilst keeping the signal processing complexity low [12]-[14]. Basically, QAM-FBMC use different prototype filters to odd- and even-numbered subcarriers to maintain complex domain orthogonality. QAM-FBCM is shown to be able to remove intrinsic interference and thus the implementation of channel estimation and MIMO is straightforward.

Another important component is modulation. QAM is envisioned to be adopted in 5G as it provides good Euclidian distance properties and it is easy to demodulate. Instead of the conventional QAM modulation, FQAM as a combination of QAM and Frequency Shift Keying (FSK) has been proposed in conjunction with OFDMA and it has demonstrated a significant performance gain especially for the cell-edge users [15]. With FQAM, the statistical distribution of inter-cell interference is likely to be non-Gaussian, especially for cell-edge users. It is proved that the worst-case additive noise in wireless networks with respect to the channel capacity has a Gaussian distribution. From this observation, one can expect that the channel capacity can be increased by FQAM which makes inter-cell interference non-Gaussian. In addition, the fact that only a small number of sub-carriers are activated in FQAM results in low PAPR so that the energy efficiency is improved, thereby makes it a suitable solution for MTC devices with stringent energy consumption requirements.

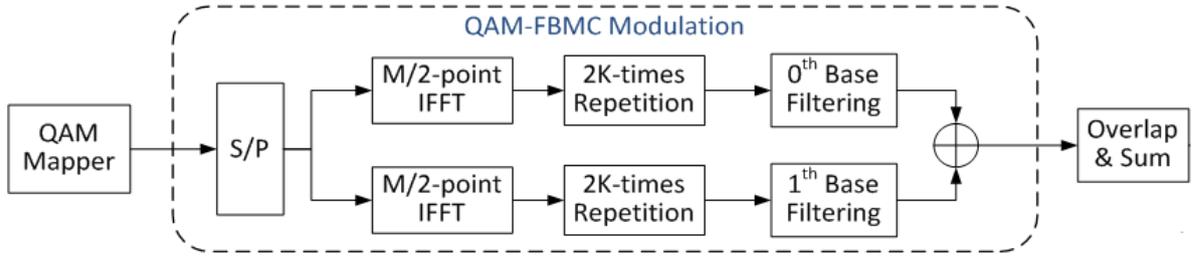

Figure 1 Conventional QAM-FBMC

However, trying to harness the similar benefits by applying FQAM in an FBMC-based system through direct integration of the two might not be able to fulfill the key objectives for modulation and waveform design for MTC devices: 1) low complexity, e.g., computation complexity and implementation complexity, 2) low self-interference, 3) good spectrum confinement, i.e., low out-of-band (OOB) emission, and 4) good frequency localization, i.e., small side-lobe. In this regard, direct implementation of FQAM in current QAM-FBMC may cause following problems:

- different prototype filters are needed for odd- and even-numbered subcarriers, which increases the implementation complexity;
- when the spectrum confinement and frequency localization are good, there is strong self-interference in the complex domain;
- when self-interference is small, spectrum is not well confined.

In this paper, we propose a new method to integrate FQAM and FBCM to achieve four main objectives at the expense of only a minor transmission rate loss. The rest of the paper is organized as follows. In section II of this paper, a system model is provided. Then the proposed solutions are introduced in Section III. Section IV is devoted to analyzing the performance of the proposed approach in terms of complexity, self-interference, spectrum confinement and frequency localization. Some numerical results are presented in section V and the paper is concluded in section VI.

## II. SYSTEM MODEL

The conventional QAM-FBMC system separates adjacent subcarriers with $B$ filter-banks keeping near orthogonality in complex domain. The transmitted signal can be expressed as

$$x(n) = \sum_{k=-\infty}^{+\infty} \sum_{b=0}^{B-1} p_b[n-kM] \left( \sum_{s=0}^{\frac{M}{B}-1} D_{b,s}[k] e^{j\frac{2\pi n}{M/B}s} \right) \quad (1)$$

where $D_{b,s}[k]$ is the complex data symbol on the $(sB+b)$-th subcarrier in the $k$-th symbol, $M$ is the total number of subcarriers and $p_b[n]$ is the $b$-th prototype filter. The time domain filter coefficients $p_b[n]$ is given by $\{n|0\leq n<LM\}$, where $L$ is called an overlapping factor. With $L\geq 2$, the QAM-FBMC symbols have duration $LM$ and thus partially overlap. The QAM-FBMC system can be efficiently implemented using $M/B$-point inverse fast Fourier transform (IFFT), $BL$-times repetition, and time domain base filtering as shown in Figure 1 [13] for the case of $L=2$.

FQAM is a combination of Frequency Shift Keying (FSK) and Quadrature Amplitude Modulation (QAM) [15]. A ($M_F$, $M_Q$) F-QAM symbol carries $\log_2 M_F + \log_2 M_Q$ bits, where $\log_2 M_F$ bits represent one of the frequency tones, and $\log_2 M_Q$ represent one of the QAM symbols as shown in Figure 2.

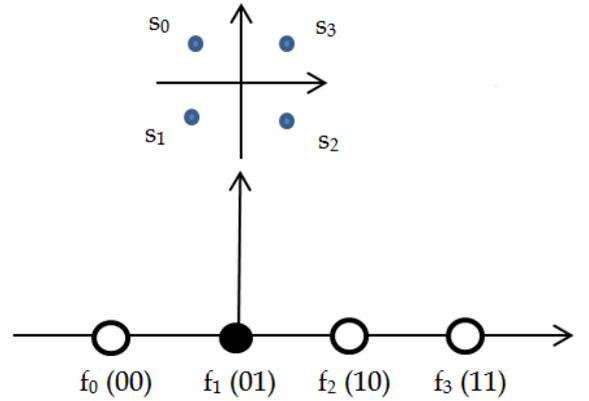

Figure 2 FQAM

Three design options are proposed in the current QAM-FBMC systems [12]-[13], where different prototype filters are used for odd- and even-numbered subcarriers, i.e., $B=2$. As a simple extension of the QAM implementations, i.e. without taking other issues (such as spectrum confinement and complexity into consideration), FQAM can be implemented in these solutions as

$$x(n) = \sum_{k=-\infty}^{+\infty} \sum_{b=0}^{B-1} p_b[n-kM] \left( \sum_{s=0}^{\frac{M}{B}-1} \delta_{f_{b,s},l_{b,s}} D_{b,s}[k] e^{j\frac{2\pi n}{M/B}s} \right) \quad (2)$$

where $f_{b,s} \in \{0, 1, …, M_F-1\}$ is the frequency-tone index based on FSK modulation for the $(sB+b)$-th subcarrier, $l_{b,s} = (sB+b)$ modulo $M_F$, and $\delta_{f,l}$ is the Kronecker delta function defined as

$$\delta_{f,l} = \begin{cases} 1, & f = l, \\ 0, & f \ne l. \end{cases} \quad (3)$$

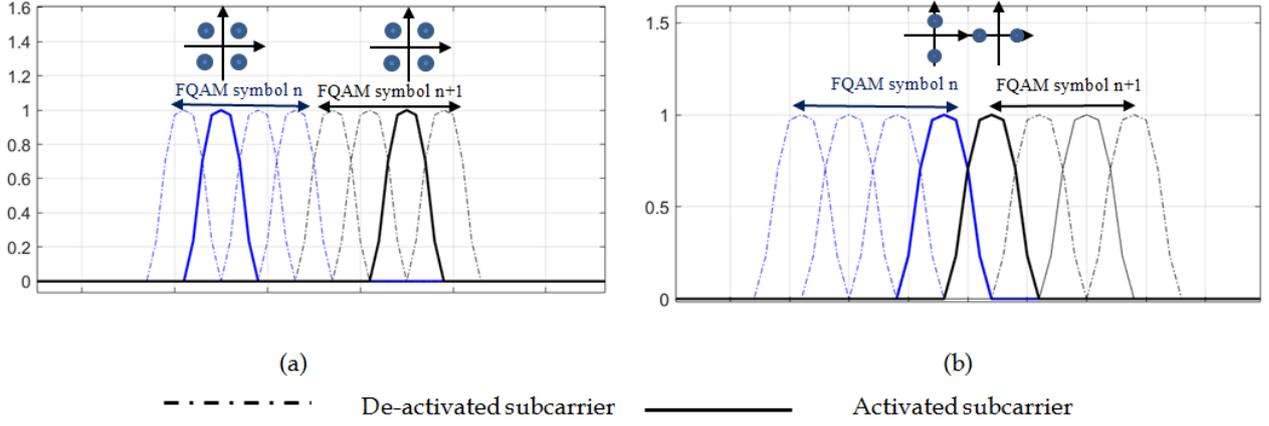

Figure 3 Self-interference in frequency domain for scheme 1

### III. PROPOSED APPROACH

In contrast to the current solution, we propose to use the same prototype filters for all subcarriers, meaning that FQAM can be implemented as

$$x(n) = \sum_{k=-\infty}^{+\infty} \sum_{m=0}^{M-1} \delta_{f_m,l_m} p[n-kM] D_m[k] e^{j\frac{2\pi n}{M}m} \quad (4)$$

where $f_m \in \{0, 1, …, M_F-1\}$ is the frequency-tone index based on FSK modulation for the $m$-th subcarrier, $l_m = m$ modulo $M_F$.

The use of the same prototype filter for odd- and even-numbered subcarriers significantly reduces implementation complexity. The prototype filter should be carefully designed to provide good spectrum confinement and therefore PHYDYAS filter is employed [9]. In traditional QAM-FBMC, the level of self-interference mainly depends on the filter shape and spacing between two adjacent subcarriers, denoted as $\Delta f$. In FQAM-FBMC, only one out of $M_F$ subcarriers is activated to carry a QAM symbol and the rest of the subcarriers are deactivated. Therefore, the level of self-interference depends on the spacing between two active subcarriers carrying QAM symbols, which is a multiple of $\Delta f$, anywhere between $1 \times \Delta f$ and $(M_F-1) \times \Delta f$. In this regard, two approaches are proposed to eliminate self-interference.

#### A. Scheme 1: FQAM(ASK)-FBMC

With PHYDYAS filter, we know that as long as the spacing between two adjacent active subcarriers is larger than $2\Delta f$, the self-interference could be extremely small [9]. As only one out of $M_F$ subcarriers is activated to carry a QAM symbol and the rest of the subcarriers are deactivated, it follows that as long as $f_m \ne 0$ or $M_F-1$, i.e., the subcarriers on the two edges are not activated, no self-interference will be generated from FQAM symbol $n$ to adjacent FQAM symbol $n\pm1$ as shown in Figure 3 (a), where the spacing between two activated subcarriers is $5\Delta f$. Therefore, QAM can be used in conjunction with FSK to form a FQAM symbol.

On the contrary, if the subcarriers on the edge are activated as shown in Figure 3 (b), strong complex domain self-interference will occur between two FQAM symbols demonstrated by the overlapping part of two activated edge subcarriers. In such a case, complex domain self-interference cannot be eliminated. Therefore, ASK modulation is proposed to be applied in either real or imaginary domain only on the edge subcarriers that might cause self-interference to ensure complex domain orthogonality and the modulation is therefore FASK. When ASK is applied, the transmission rate in terms of bits/symbol is reduced (since we use ASK instead of QAM) but the rate loss could be small depending on FSK modulation order.

#### B. Scheme 2: Opportunistic FQAM-FBMC

In the scheme 1, ASK is applied as long as an active subcarrier is possible to generate self-interference. In scheme 2, the spacing between adjacent active subcarriers are taken into consideration and QAM or ASK modulation is opportunistically applied. By doing this, the rate loss can be further reduced.

Instead of only transmitting in real/imaginary domain, we propose that after FSK modulation, if the subcarrier at either edge of one FQAM symbol is activated, its adjacent FQAM symbol will be checked. There is a possibility that the adjacent FQAM symbol does not activate the subcarrier that may cause self-interference as shown in Figure 4. In such a case, QAM symbols can still be applied even when $f_m=0$ or $M_F$, i.e., the edge subcarrier is activated. Only when two adjacent subcarriers are activated at the same time, ASK modulation in real/imaginary domain is employed as in scheme 1.

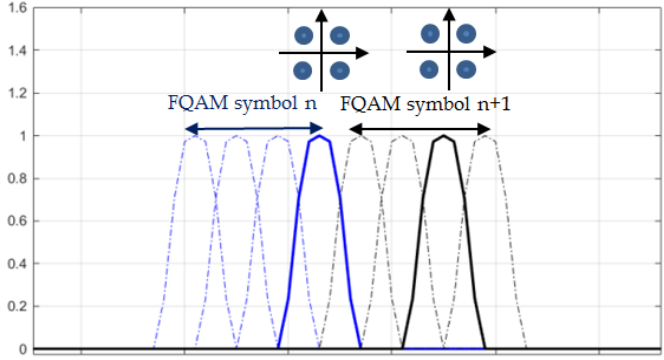

Figure 4 Opportunistic FQAM-FBMC

It should be noted that when the new schemes are applied, the difference between eq. (2) and (4) is $D_m[k]$ in (4) could be either QAM modulated complex signal or ASK modulated real/imaginary signal depending on the level of self-interference generated by the active subcarriers. In addition, the demodulation of opportunistic FQAM-FBMC should take into account the constraints imposed by the modulation.

*C. Rate Loss Analysis*

In scheme 1, FSK always carries $\log_2(M_F)$ bits. For edge subcarriers, ASK carries $1/2\log_2(M_Q)$ and the incidence of applying ASK is $2/M_F$, giving an average of $\log_2(M_Q)/M_F$ bits in total for edge subcarriers. For non-edge subcarriers, QAM carries $\log_2(M_Q)$ and the incidence of such a case is $(M_F-2)/M_F$, so that total $(M_F-2)/M_F*\log_2(M_Q)$ bits can be conveyed on average. Summing the two together, the average transmission rate, in bits per symbol, is given as

$$R = \log_2(M_F) + \frac{M_F - 1}{M_F} \log_2(M_Q) \quad (5)$$

The transmission rate will be reduced due to replacing QAM with ASK but the rate loss could be small with increased FSK modulation order. The reduction is

$$R_\Delta = \frac{1}{M_F} \log_2(M_Q) \quad (6)$$

For example, for the proposed FQAM-FBMC design, when we have $(M_F,M_Q)=(4,4)$, the rate of the proposed scheme is 3.5 bit/symbol with 12.5% rate loss compared with normal rate 4 bit/symbol. When we have $(M_F,M_Q)=(8,4)$, the rate of the proposed scheme is 4.75 bit/symbol with only 5% rate loss compared with normal rate 5 bit/symbol.

With the opportunistic approach, the transmission rate loss can be further reduced. Using similar analysis as in the first scheme, the transmission rate is given as

$$R = \log_2(M_F) + \frac{M_F^2 - 1}{M_F^2} \log_2(M_Q) \quad (7)$$

The transmission rate loss is reduced to

$$R_\Delta = \frac{1}{M_F^2} \log_2(M_Q) \quad (8)$$

For $(M_F,M_Q)=(4,4)$, the average rate of the proposed scheme is 3.875 bit/symbol with 3.13% rate loss compared with normal rate 4 bit/symbol. When we have $(M_F,M_Q)=(8,4)$, the average rate of the proposed scheme is 4.97 bit/symbol with only 0.7% rate loss compared with normal rate 5 bit/symbol.

## IV. PERFORMANCE COMPARISONS

There are three existing QAM-FBMC solutions with different prototype filter design:

• Solution 1: PHYDYAS filter and its block interleaved variant are used for odd- and even-numbered subcarriers, respectively [12].

• Solution 2: Type-I filter is used for both odd- and even-numbered subcarriers [13].

• Solution 3: Two different Type-II filters are used for odd- and even-numbered subcarriers, respectively [14].

We will compare these state-of-the-art QAM-FBMC based solutions with our proposed FQAM-FBMC scheme in terms of self-interference, spectrum confinement and complexity in this section.

*A. Self-interference Comparisons*

The signal power and the total self-interference power on $(b, s)$-th subcarrier can be calculated as

$$P_s = \left|p_{(b,s),(b,s)}[0]\right|^2$$
$$P_i = \sum_{(b',s')\neq(b,s)} \sum_{k=-(L-1)}^{L} \left|p_{(b,s),(b',s')}[kM]\right|^2 - \left|p_{(b,s),(b,s)}[0]\right|^2 \quad (9)$$

respectively, and the self-signal-to-interference ratio (self-SIR) can define as

$$\gamma_{self} = 10\log_{10} \frac{P_s}{P_i} \text{ (dB)} \quad (10)$$

Conventional solution 2 and 3 only achieve around 15 and 20dB self-SIR based on (9) and (10) because of the strong self-interference. With relatively low self-SIR, only low order modulation schemes can be applied, e.g., 4QAM and 16 QAM. On the contrary, conventional solution 1 can achieve more than 60dB self-SIR and therefore high order modulation, e.g., 64QAM, can be applied. Since it is possible to achieve more than 60dB self-SIR based on (9) and (10), in the proposed scheme the self-interference can be very small, even negligible. This is due to the fast fall-off rate provided by PHYDYAS filter.

## B. Spectrum Confinement and Frequency Localization Comparisons

In the first solution, PHYDYAS filter is used for odd-numbered subcarriers and a block interleaved PHYDYAS filter is used for even-numbered subcarriers for the sake of complex domain orthogonality. However, the block interleaving will severely undermine the spectrum confinement performance as shown in Figure 5.

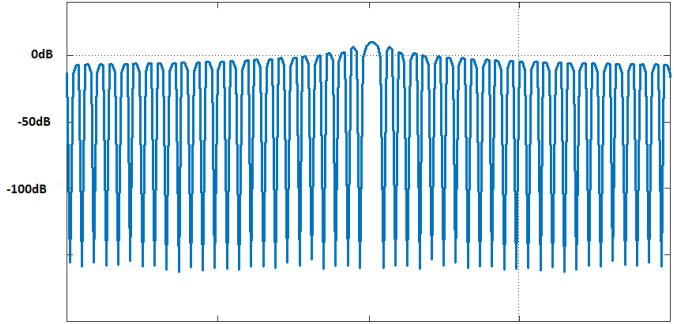

Figure 5 PSD for the odd-numbered subcarriers

The high side-lobe lead to two consequences: 1) high inter-subcarrier interference in asynchronous scenario and thus imposes very high requirement for synchronization; 2) more guard band at two edges of the entire spectrum, which will cause low spectral efficiency. Both have negative impacts on the implementation of MTC devices.

The second solution uses a prototype filter with PSD similar to that of PHYDYAS filter as shown in Figure 6 and the same filter is used for both even- and odd-numbered subcarriers. Thus its spectrum confinement performance is similar with the proposed scheme.

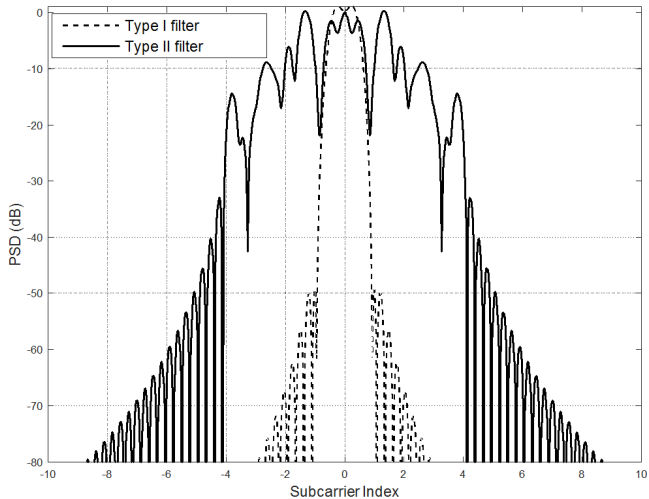

Figure 6 PSD of Type-I and Type-II filters

The PSD of Type-II filter in solution 3 is shown in Figure 6 as well. Clearly, it has a larger side-lobe and the spectrum is not confined as well as Type-I filter and the proposed scheme.

## C. Complexity

From the complexity perspective, both solution 1 and solution 3 use two different prototype filters, which clearly increases the implementation complexity compared to the proposed scheme with only one prototype filter. Therefore, the computation complexity of solution 1 and 3 is higher than the proposed scheme.

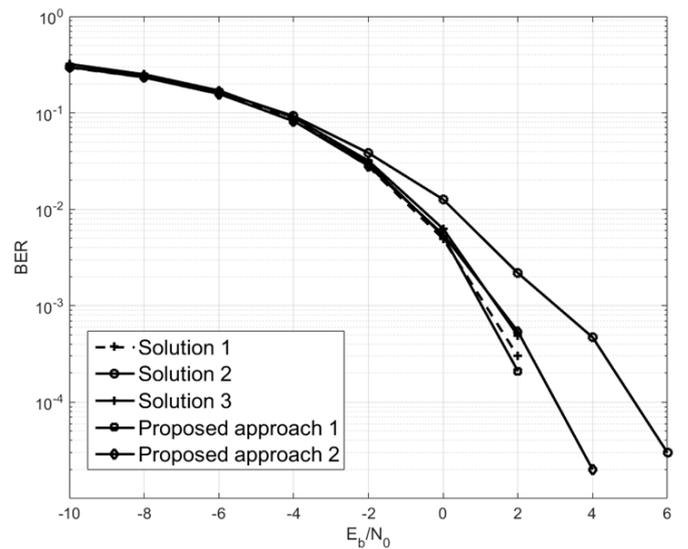

Figure 7 AWGN Channel

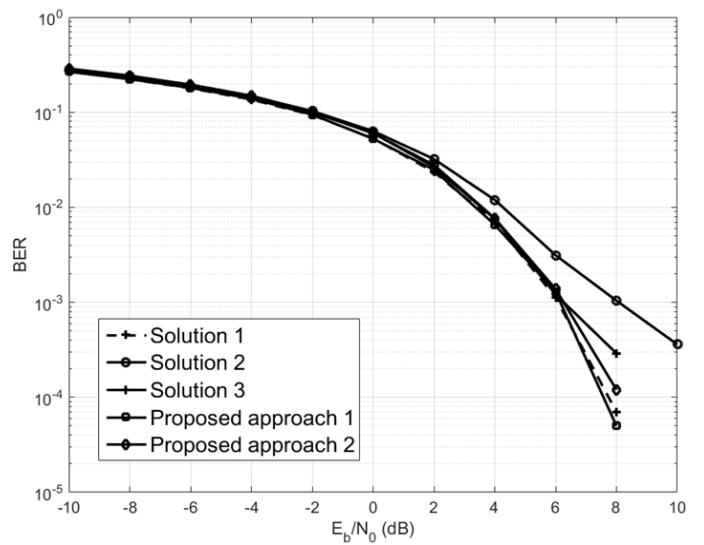

Figure 8 EVA Channel

## V. NUMERICAL RESULTS

In this section, we present numerical results to compare the performance of the proposed approaches with conventional ones. Here we assume $M$=100, $L$=4 and zero forcing (ZF) equalization is used for non-Gaussian channel. Uncoded bit error rate (BER) results for both AWGN channel (given as an initial study, applicable to conventional channels with static or low-speed users) and 3GPP Extended Vehicular A (EVA) channel with speed 50 km/h (provided to cover more realistic cases and high speeds) are presented as shown in Figure 7 and 8 [16]. The modulation order of FQAM is $(M_F, M_Q)$=(4,4), which is used to for all approaches.

As can be seen from the figures, the BER performance of the conventional solutions is determined by self-SIR. Solution with the highest self-SIR (over 60 dB), i.e., solution 1 achieves the best performance and performance of solution 2 and 3 is degraded because of the lower self-SIR (20 and 15 dB). The proposed approach 1, where ASK is used to avoid self-interference for the edge subcarriers, slightly outperforms the best conventional solution because the self-interference can be completely removed. The proposed approach 2 has slightly higher BER than the conventional solution 1 and achieves the same BER as conventional solution 2. Both new approaches suffer certain level of rate loss but the loss of the second approach is negligible. More importantly, both new approaches have lower complexity and better spectrum confinement.

Finally, a qualitative comparison of the above schemes is presented in Table-I.

Table-I Comparisons

|  | Solution 1 | Solution 2 | Solution 3 | New schemes |
|---|---|---|---|---|
| Complexity | Medium | Low | High | Low |
| Self-interference | Low | High | Medium | Low |
| Spectrum confinement | Bad | Good | Medium | Good |
| Frequency localization | Bad | Good | Medium | Good |

## VI. CONCLUSIONS

There are some major requirements for the design of air-interfaces for MTC devices: 1) low complexity, 2) low energy consumption and 3) better spectrum confinement considering MTC devices might need to operate in an asynchronous manner. In this paper, we propose novel approaches to efficiently integrate FQAM and FBMC for MTC devices to fulfill these objectives. The PAPR of FQAM is lower because only 1 subcarrier is activated out of $M_F$ subcarriers; this additionally reduces energy consumption. Whereas two different prototype filters are used for odd- and even-numbered subcarriers in conventional QAM-FBMC, only one prototype filter is employed in the new approaches to reduce implementation complexity. In addition, the proposed solutions also achieve low self-interference by utilizing the unique characteristics of FQAM. Better spectrum confinement is achieved using the PHYDAS prototype filter.

ACKNOWLEDGEMENTS

This work has been performed in the framework of the H2020 project METIS-II co-funded by the EU. The views expressed are those of the authors and do not necessarily represent the project. The consortium is not liable for any use that may be made of any of the information contained therein.